\documentstyle[prb,aps,amsfonts,amssymb,multicol,epsf]{revtex}

\begin{document}
\draft

\title
{\bf Velocity-force characteristics of a driven interface in a disordered
medium}

\author{M.\ M\"uller$^{\,1}$, D.A.\ Gorokhov$^{\,1,2}$, and G.\ Blatter$^{\,1}$}

\address{$^{1\,}$Theoretische Physik, ETH-H\"onggerberg, CH-8093 Z\"urich, Switzerland}
\address{$^{2\,}$Department of Physics, Harvard University, Cambridge, MA 02138, USA}

\date{\today}
\maketitle

\begin{abstract}

Using a dynamic functional renormalization group treatment of driven elastic
interfaces in a disordered medium, we investigate several aspects of the
creep-type motion induced by external forces below the depinning threshold
$f_c$: {\it i)} We show that in the experimentally important regime of forces
slightly below $f_c$ the velocity obeys an Arrhenius-type  law
$v\sim\exp[-U(f)/T]$ with an effective energy barrier $U(f)\propto (f_{c}-f)$
vanishing linearly when $f$ approaches the thres\-hold $f_c$. {\it ii)} Thermal
fluctuations soften the pinning landscape at high temperatures. Determining the
corresponding velocity-force characteristics at low driving forces for internal
dimensions $d=1,2$ (strings and interfaces) we find a particular non-Arrhenius
type creep $v\sim \exp[-(f_c(T)/f)^{\mu}]$ involving the reduced threshold force
$f_c(T)$ alone. For $d=3$ we obtain a similar $v$--$f$ characteristic which is,
however, non-universal and depends explicitly on the microscopic cutoff.
\end{abstract}

\pacs{PACS numbers: 05.20.-y, 64.60.Cn, 74.60.Ge, 82.65.Dp}

\begin{multicols}{2}

\section{Introduction}

The influence of disorder on the static and dynamic properties of elastic
systems has been intensely studied in recent
years\cite{HalpinHealy,ScheidlNattermann}. Various physical systems including
flux lattices in superconductors\cite{Blatter}, domain walls in
magnets\cite{Lemerle,Forgacs}, and charge density waves in solids\cite{Gruner}
significantly change their properties upon introducing even a small amount of
disorder. Subject to a disorder landscape, these systems transform to a glassy
state characterized by a nontrivial scaling of the displacement correlation
functions\cite{HuseHenley} and a vanishing linear response to external driving
forces\cite{VinokurIoffe,MPAFisher,Feigelman,Nattermann1990}, e.g., the current
induced Lorentz force acting on vortices or the magnetic field driving the
domain walls in magnets. The determination of the velocity-force characteristics
of a driven elastic manifold subject to a disorder landscape is a challenging
problem: while the behavior at small distances and large drives is amenable to
perturbation theory, the most interesting long distance/weak drive regime can
only be attacked via non-perturbative methods. In this paper, we consider some
aspects of the creep-type dissipative motion of a driven elastic interface with
$d$ internal dimensions, moving along one transverse direction in a disorder
landscape ($d+1$-dimensional random manifold problem).

\begin{figure}
\centerline{\epsfxsize=8.5cm \epsfbox{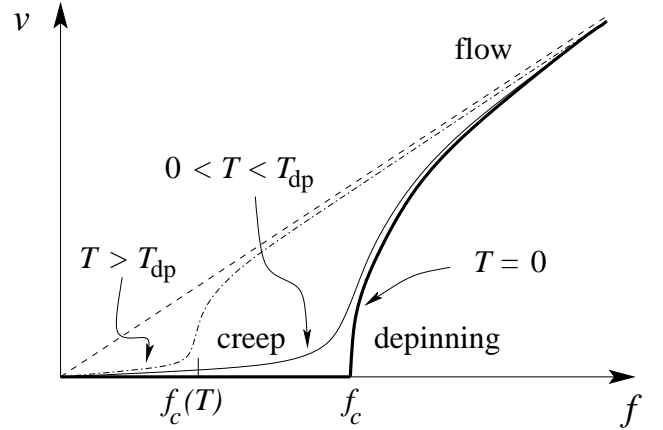}} \narrowtext\vspace{5mm}
\caption{Velocity-force characteristic of a driven interface. The thick solid
line is the zero-temperature result with a thres\-hold force $f_{c}$ below which
the velocity vanishes. Beyond $f_{c}$ the velocity first rises following the
scaling law $v \propto (f-f_{c})^{\beta}$ and then crosses over to the linear
dissipative regime with $v \propto f$. The thin line shows the behavior at
finite but low temperatures with a creep regime at low forces $f \ll f_c$,
$v\propto\exp[-(U_{c}/T) {(f_c/f)}^{\mu}]$. Close to threshold we find that the
creep barriers scale linearly in $f$, $v \propto \exp[-({U_c}/T) (1-f/f_{c})]$.
At high temperatures (dash-dotted line) thermal fluctuations become particularly
important in dimensions $d=1,2$; the threshold force $f_{c}(T)$ is strongly
reduced by thermal fluctuations and we find a non-Arrhenius glassy response at
small drive with $v\propto\exp[-(f_{c}(T)/f)^{\mu}]$ determined by the
renormalized critical force alone.} \label{F1}
\end{figure}

Depending on the value of the temperature $T$ and the external force $f$ several
regimes can be distinguished (see Fig.\ \ref{F1}): At $T=0$, the velocity $v$ is
zero as long as $f$ does not exceed the critical force $f_{c}$, whereas for $f >
f_{c}$ the system starts moving, $v(f)\ne 0$. In particular, one finds
$v(f)\propto {(f-f_{c})}^{\beta}$ near the threshold (the depinning transition),
with a nontrivial critical exponent
$\beta$\cite{Nattermann,NarayanFisher,Leschhorn,Chauve}. For large drives $f \gg
f_{c}$ the disorder becomes irrelevant and the velocity-force characteristic
turns linear, $v \sim f/\eta$, with $\eta$ the friction coefficient
characteristic of the dissipative dynamics.

At finite temperatures $T>0$, thermal fluctuations induce a creep-type motion
resulting in an exponentially small but finite velocity even below threshold $f
< f_{c}$ (see Fig.\ \ref{F1}). At small drives $f \rightarrow 0$ an
Arrhenius-type law $v(f) \propto \exp\left\{-U(f)/T\right\}$ holds, with a
diverging activation barrier $U(f \rightarrow 0) \rightarrow \infty$ (glassy
response).
Close to threshold $f \sim f_c$ one may distinguish two interesting regimes:
{\it i)} fixing the force $f$ at its critical value $f = f_c$, thermal
fluctuations smooth the depinning transition and the velocity $v_{f_c}(T)
\propto T^{1/\delta}$ is expected to scale as a power of
temperature\cite{Fisher85,Middleton1}. {\it ii)} fixing the temperature $T$ and
increasing the force towards threshold $f \rightarrow f_{c}$, a creep-type
response is expected with a vanishing activation barrier $U(f\rightarrow f_c)
\propto (f_c-f)^\alpha \rightarrow 0$; here we are interested in the second
situation.

The scaling theory of
creep\cite{VinokurIoffe,MPAFisher,Feigelman,Nattermann1990} predicts that
$U(f\rightarrow 0) \propto U_{c}{\left(f_{c}/f\right)}^{\mu}$, with a
characteristic energy scale $U_{c}$ set by the disorder landscape. On the other
hand, when the force $f$ approaches $f_{c}$ from below one expects that the
barrier behaves like
\begin{equation}
U(f)\propto {(f_{c}-f)}^{\alpha}, \label{barrier}
\end{equation}
with $\alpha$ an exponent depending on the dimensionality of the space and the
elastic manifold. The parameter $\alpha$ determines the relaxation of
magnetization in superconductors at currents close to the critical one. In his
original description of magnetic relaxation, Anderson\cite{Anderson} assumed
that $\alpha =1$ and explained the famous logarithmic decay of the magnetic
field trapped inside a superconductor\cite{Kim}. Note that the regime $f \simeq
f_c$ is an experimentally important one: the observation of the system response
at small driving forces $f \ll f_c$ involves long relaxation times $\propto
\exp[(U_{c}/T)(f_{c}/f)^{\mu}]$ and, hence, this regime is more difficult to
access experimentally. More quantitatively, the maximal creep barrier $U$ that
can be observed after a waiting time $t$ is given by $U(t) \approx T \ln
(1+t/t_0)$ with $t_0$ a characteristic time scale involving details of the
critical state \cite{Geshkenbein,Blatter}. For the vortex creep problem this
time scale typically is of order $10^{-6}$\,s and thus the experimentally
attainable value of $U/T$ is limited to factors $\sim 30$--40.

From a theoretical point of view the calculation of the barrier exponent
$\alpha$ in (\ref{barrier}) near criticality still remains a problem. In fact,
one may expect that critical fluctuations of the manifold near the threshold
will affect the creep motion. In this paper we study the behavior of $U(f)$ near
the threshold using dynamical renormalization group theory\cite{Chauve} and show
that if the pinning of the manifold is due to a short-range correlated random
potential (e.g., due to point-like impurities) the effective barrier behaves as
\begin{equation}
U(f)\simeq U_{c}\left(1-f/f_c\right), \qquad f \rightarrow f_{c},
\label{critbarr}
\end{equation}
with $U_{c}$ a characteristic energy scale set by the disorder landscape. This
result is independent of the dimensionality of the manifold and confirms the
original assumption of Anderson\cite{Anderson}.

In addition, we investigate creep at high temperatures, again using the
dynamical functional renormalization group technique. In this case, the
dimensionality of the manifold is particularly important: It is well-known that
the mean thermal displacement $\langle u^2\rangle_{\rm th}$ of a manifold with
internal dimension $d \ge 3$ is bounded, the maximum displacement depending on
the microscopic short-scale cut-off of the elastic system. Strings and surfaces
($d=1,\,2$), however, exhibit thermal fluctuations $\langle [{\bf u}({\bf
z})-{\bf u}({\bf 0})]^2\rangle$ which grow unboundedly with separation $z$. At
high temperatures, the manifold probes an effective disorder landscape averaged
over thermal displacements which are only bounded through the disorder-induced
pinning at large scales, resulting in a strongly reduced disorder strength. In
particular, the critical force $f_c(T)$ is found to decrease as a power law with
increasing temperature, $f_c(T) \sim f_c (T_{\rm dp}/T)^\kappa$ with $\kappa =
7$ (2) in $d=1$ (2) dimensions (see Fig.\ \ref{F1}). The characteristic
temperature $T_{\rm dp}$ determining the crossover from the low to the high
temperature regime is given by the bare disorder energy scale $U_c$, $T_{\rm dp}
= U_c$. For the effective barrier in $d=1,\,2$ we find
\begin{equation}
U(f) \simeq T \left [{\left(f_c(T)/f\right)}^{\mu}-1\right], \label{result}
\end{equation}
depending only on the renormalized pinning force $f_c(T)$, confirming the
results obtained previously via scaling estimates, see Ref.\
\onlinecite{Blatter}. Similarly to the low $T$ creep, the exponent $\mu
=(d+2\zeta -2)/(2-\zeta)$ again is determined through the static roughness
exponent $\zeta$. The large thermal fluctuations modify the characteristic
energy scale $U_c$ of the problem to $U_c \rightarrow T$, leading to a peculiar
non-Arrhenius form for the $v-f$-characteristic in the high $T$ creep regime.
For $d=3$ the velocity--force characteristic also takes a non-Arrhenius form but
with an exponent additionally modified through the temperature dependence of the
creep barrier $U_c(T) \gg T$. Note, however, that Eqs.~(\ref{critbarr}) and
(\ref{result}) make sense only if the temperature $T$ is still small enough to
produce an exponent $U(f)/T \gg 1$.

In Sec.\ II below we will first analyze creep near the critical force and derive
Eq.~(\ref{critbarr}) while section~\ref{temperaturecreep} is devoted to the
study of creep at high temperatures with a derivation of the result
Eq.~(\ref{result}) for the creep barrier.

\section{Creep near threshold}\label{threshold}

In the following we will concentrate on the case of $(d+1)$-dimensional elastic
media, with $d$  internal dimensions and one single transverse direction.
Typical realizations are strings confined to a plane (a $(1+1)$-dimensional
manifold) or two-dimensional membranes embedded in three-dimensional space (a
$(2+1)$-dimensional manifold). These models describe domain walls in thin film
and bulk random magnets, for example. The motion of the elastic manifold is
governed by the equation
\begin{equation}
\label{eqmot} \eta \partial_t u=c \nabla_{{\bf z}}^2 u+f_{\rm pin}(u,{\bf
z})+\zeta({\bf z},t)+f,
\end{equation}
where the friction and external driving forces are given by $\eta \partial_t u$
and $f$ respectively, and the additional forces acting on the manifold are those
due to elasticity $c\nabla_{{\bf z}}^2 u$, pinning $f_{\rm pin}(u,{\bf z})$, and
thermal fluctuations $\zeta({\bf z},t)$; $\eta$ and $c$ denote the viscosity and
the elasticity per unit volume. We assume that the pinning force is a Gaussian
random variable with zero mean and a correlator $\langle f(u, {\bf z})
f(u^{\prime}, {\bf z}^{\prime})\rangle = \Delta (u-u^{\prime})\delta^d ({\bf
z}-{\bf z}^{\prime})$ of width $\xi$, the typical length scale of the disorder
landscape. The statistics of the stochastic force $\zeta ({\bf z}, t)$ is
Gaussian as well and the correlator is related to the viscosity $\eta$ and the
temperature $T$ via $\langle \zeta ({\bf z}, t)\zeta ({\bf z}^{\prime},
t^{\prime})\rangle = 2\eta T\delta^d ({\bf z}-{\bf z}^{\prime})\delta
(t-t^{\prime})$.

The calculation of the average velocity $v=\langle \partial_{t}u\rangle$ as a
function of $f$ and $T$ is a difficult problem in the creep regime $f<f_{c}$
since most of the time the manifold is pinned by the random potential and only
rarely a strong thermal fluctuation will drive it into a neighboring metastable
state. Obviously, this type of motion cannot be described perturbatively.
However, it can rigorously be proven that the velocity-force characteristic is
unique\cite{Middleton}.

A powerful method to study random elastic manifolds is the functional
renormalization group (FRG)\cite{DSFisher,BalentsFisher,GiamarchiLeDoussal} with
various extensions dealing with finite temperature\cite{Balents} and
velocity\cite{Nattermann,NarayanFisher,Leschhorn,Chauve}. For dimensionalities
$d$ of the manifold larger than four, the effect of disorder can be taken into
account perturbatively, whereas in less than four dimensions, an
$\epsilon$-expansion allows to study the properties of the system at small
$\epsilon = 4-d$. The FRG has provided numerous results in the investigation of
static and dynamic properties of elastic manifolds: The static wandering
exponent\cite{DSFisher} $\zeta$ as well as the dynamic exponent\cite{Leschhorn}
$z$ have been determined for different types of disorder. Furthermore, the
depinning transition at $T=0$ has been analyzed and the critical exponent
$\beta$ in the depinning law $v\propto (f-f_{c})^{\beta}$ has been
calculated\cite{Nattermann,NarayanFisher,Leschhorn,Chauve}.

The dynamical extension of the FRG by Chauve {\it et al.}\cite{Chauve} allows to
investigate the creep regime and confirms the creep law $U(f)\propto U_{c}
(f_{c}/f)^{\mu}$ derived earlier via scaling arguments. In addition, it turns
out that the method allows for the determination of different characteristics of
the manifold's dynamics without additional physical assumptions (c.f.
Refs.~\cite{Nattermann,NarayanFisher,Leschhorn}). The dynamical FRG starts from
a Martin-Siggia-Rose action\cite{MSR} obtained from the equation of motion
Eq.~(\ref{eqmot}) and proceeds with the elimination of large momentum
fluctuations. Thereby the parameters entering Eq.~(\ref{eqmot}) are renormalized
and characterize a system for which disorder is less and less relevant. Finally,
the flow is cut off at a scale where the effect of disorder can be taken into
account perturbatively.

Our starting point is the system of equations derived in Ref.\
\onlinecite{Chauve} describing the renormalization of the parameters entering
Eq.~(\ref{eqmot}), to lowest nontrivial order
\begin{eqnarray}
\partial_l\tilde{\Delta}_l(u)
 &=&(\epsilon\!-\!2\zeta)\tilde{\Delta}_l(u)\!+\zeta u\tilde{\Delta}_l'(u)
      \!+\tilde{T}_l\tilde{\Delta}_l''(u)
      \!+\!\!\int_{s>0,s'>0} \!\!\!\!\!\!\!\!\!\!\!\!\!\!\!\!\! e^{-s-s'}\nonumber\\
 &\times& \big[\tilde{\Delta}_l''(u) [\tilde{\Delta}_l(\lambda_l(s'-s))
% \nonumber \\
% &&\qquad\qquad\qquad
   -\tilde{\Delta}_l(u+\lambda_l(s'-s))]\nonumber \\
 &&-\tilde{\Delta}_l'(u-\lambda_l s')
    \tilde{\Delta}_l'(u+\lambda_l s)
 \label{eq:RGcreep1} \\
 &&+\tilde{\Delta}_l'(\lambda_l(s'+s))
                       [\tilde{\Delta}_l'(u-\lambda_l s')-
                        \tilde{\Delta}_l'(u+\lambda_l s)]\big],\nonumber\\
 \label{eq:RGcreep2}
 \partial_l \ln\lambda_l
 &=& 2-\zeta-\int_{s>0} e^{-s} s \tilde{\Delta}_l''(\lambda_l s),\\
 \label{eq:RGcreep4}
 \partial_l \tilde{{f}}_l
 &=& (2-\zeta)\tilde{{f}}_l + c\Lambda^2 \int_{s>0} e^{-s}
              \tilde{\Delta}_l'(\lambda_l s),\\
 \label{eq:RGcreep3}
 \partial_l \ln\tilde{T}_l
 &=& - \theta + \int_{s>0}e^{-s}\lambda_l s \tilde{\Delta}_l'''(\lambda_l s),
\end{eqnarray}
with ${\tilde\Delta}_{l}(u) = \left (A_{d}\Lambda^{d-4} /c^2\right
)\Delta_{l}(u)$, $\lambda_l = (\eta v)_{l}/c\Lambda^{2}$, ${\tilde T}_{l} =
A_{d}\Lambda^{d-2}T_{l}/c$, and ${\tilde f}_{l} = f_{l}-(\eta v)_{l}$. The
exponents $\zeta$ and $\theta =d-2+2\zeta$ describe the scaling of the roughness
and the energy, respectively. $A_d$ is the surface of the unit sphere in $d$
dimensions and $\Lambda$ denotes the short-scale cut-off of the theory. Note the
important effect of the dynamics in rendering the equations non-local on the
scale $\lambda_l$ proportional to the center of mass velocity $v$ of the
manifold. The main goal of this section is to investigate these equations in the
limit when the external force acting on the manifold is slightly below the
threshold force, $f < f_c$.

We first analyze the system of equations (\ref{eq:RGcreep1}) --
(\ref{eq:RGcreep3}) for the case of an infinitesimal velocity $v=0+$ and
concentrate on low temperatures. Eqs.~(\ref{eq:RGcreep1}) and
(\ref{eq:RGcreep3}) then reduce to the static FRG equations\cite{DSFisher},
\begin{eqnarray}
 \partial_{l}{\tilde\Delta}_{l}(u)&=&(\epsilon -2\zeta ){\tilde\Delta}_{l}(u)
  +\zeta u {\tilde\Delta}_{l}^{\prime}(u)
  +\tilde{T}_l{\tilde\Delta}_l^{\prime\prime}(u)\nonumber \\
  &&\qquad
  +{\tilde\Delta}^{\prime\prime}_{l}(u) \left({\tilde\Delta}_{l}(0)
  -{\tilde\Delta}_{l}(u)\right)-{{\tilde\Delta}_{l}^{\prime }(u)}^{2},
  \label{static}\\
  \partial_l\tilde{T}_l&=&-\theta\tilde{T}_l.
\label{Tscaling}
\end{eqnarray}
The flow takes the correlator ${\tilde \Delta}_l$ through a special point
$l_{c}\approx (1/\epsilon )\ln [\epsilon/(3|{\tilde \Delta}^{\prime\prime}_{0}
(0)|)]$, $L_c \approx \Lambda^{-1}e^{l_c} \approx \{\epsilon c^2 \xi^2/[A_d
\Delta_0(0)]\}^{1/(4-d)}$, where it becomes singular at the origin in the limit
$T \rightarrow 0$; this is easily seen from the equation
\begin{equation}
  \partial_{l}{\tilde\Delta}_{l}^{\prime\prime}(0)\approx
  \epsilon {\tilde\Delta}_{l}^{\prime\prime}(0)-
  3{\tilde\Delta}_{l}^{\prime\prime}(0)^{2},
  \label{origin}
  \end{equation}
satisfied by the second derivative of the correlator at low temperatures;
exploiting the fact that ${\tilde\Delta}_{0}^{\prime\prime}(0)<0$ we have
$|{\tilde\Delta}_l^{\prime\prime}(0)| \approx |{\tilde\Delta}_0^{\prime
\prime}(0)| e^{\epsilon l}/[1-(3 |{\tilde\Delta}_{0}^{\prime\prime}(0)|
/\epsilon) (e^{\epsilon l} - 1)]$. The curvature
${\tilde\Delta}_{l}^{\prime\prime}(0)$ diverging at $l_c$ marks the occurrence
of a non-analyticity at the origin which is reflected in the appearance of a
cusp in $\tilde{\Delta}_{l>l_c}$ at $u=0$. Although the initial correlator
${\tilde\Delta}_{0}(u)$ is usually an analytic and even function of the
coordinate $u$ with vanishing odd derivatives at the origin, the function
${\tilde\Delta}_{l>l_{c}}$  has a cusp with a nonzero slope
${\tilde\Delta}_{l}^{\prime}(0+)<0$ at the origin when $T=0$. Asymptotically,
${\tilde\Delta}_{l}(u)$ approaches a zero temperature `cuspy' fixed point
${\tilde\Delta}^{*}(u)$ describing the disordered phase with a nontrivial
roughness exponent $\zeta$. Assuming that the rough shape of the fixed point
function ${\tilde\Delta}^{*}(u)$ is assumed at the Larkin scale $l_c$ we can
easily find its characteristics: The width
\begin{equation}
\xi^*\approx \xi \exp(-\zeta l_c) \label{width}
\end{equation}
of ${\tilde\Delta}^{*}(u)$ follows from integrating the second term in
(\ref{static}). Comparing terms in (\ref{static}) at the origin $u=0$ we find
$\tilde{\Delta}^{*\prime 2}(0+) \sim \epsilon\tilde{\Delta}^{*}(0)$ and
combining this with the relation $|\tilde{\Delta}^{*\prime}(0+)| \xi^{*} \sim
\tilde{\Delta}^{*}(0)$ we find the estimates $\tilde{\Delta}^*(0) \sim \epsilon
\xi^2 \exp(-2\zeta l_c)$ and
\begin{equation}
 |\tilde{\Delta}^{*\prime}(0+)| \approx \epsilon \xi e^{-\zeta l_c}.
\label{slope}
\end{equation}

Let us then analyze the force flow (\ref{eq:RGcreep4}) in the light of these
results. The scale $l_c$ divides the flow into two distinct regimes, the Larkin
regime at small scales $l<l_c$ and the random manifold regime ($l>l_c$). For
$l<l_c$ we have ${\tilde\Delta}_{l}^{\prime} (0)={\tilde\Delta}_{l}^{\prime}
(0+)=0$ and the force ${\tilde f}_{l}$ obeys the equation $\partial_{l}{\tilde
f}_l=(2-\zeta ){\tilde f}_{l}$, i.e., ${\tilde f}_{l}=e^{(2-\zeta )l_{c}}{\tilde
f}$ grows exponentially. At the point $l=l_{c}$ the integral term on the
right-hand side of Eq.~(\ref{eq:RGcreep4}) jumps from zero to a finite value,
since the slope of the correlator at the origin does not vanish any longer. If
this contribution overcompensates the scaling term, i.e., $c\Lambda^{2}|
\Delta^{\prime}_{l_{c}} (0+)| > (2-\zeta )e^{(2-\zeta )l_{c}}f$, the force will
start renormalizing to zero while in the opposite case it will continue to
increase. This can be interpreted in the following way: if the initial value $f$
of the external force is smaller than a critical value $f_{c}$ the force will
eventually renormalize downwards and cannot move the manifold. For $f > f_{c}$
the manifold starts moving and one should take into account
Eq.~(\ref{eq:RGcreep2}) since the problem is not static any more. We therefore
come to the conclusion that for a system with a dynamics described by
Eq.~(\ref{eqmot}) there exists a finite threshold force $f_c$ at $T=0$. Using
the above condition one easily finds $f_c \approx [c\Lambda^2
|\tilde{\Delta}^{*\prime}(0+)|/(2-\zeta)] e^{-(2-\zeta)l_c} \approx
[\epsilon/(2-\zeta)] c \xi / L_c^2$ with $L_c = \Lambda^{-1} e^{l_c}$. Note that
the expression for $f_c$ coincides with the result obtained from simple scaling
estimates\cite{Blatter}.

At finite temperatures $T > 0$ it is not possible any more to define the
critical force density $f_c$ as the thres\-hold below which there is no center
of mass motion of the manifold, as thermally activated jumps lead to an average
velocity $v>0$ at any finite force $f$. Of course, at low temperatures the
velocity is exceedingly small, given that it obeys an Arrhenius-type law.
Therefore the time needed to observe this velocity might well exceed the time
scale of the experiment, i.e., from an experimental point of view the critical
force density still exists with the thres\-hold $f_c$ separating creep-type
motion from viscous flow. On the mathematical level the nonexistence of the
critical force density can be explained as follows: at $T > 0$, the slope
${\tilde\Delta}^{\prime}(0+)$ remains zero beyond $l_c$,
${\tilde\Delta}^{\prime}_{l > l_{c}}(0+) = 0$, and the renormalized force
density will continue to grow beyond the length scale $l_{c}$ even if the
initial force density $f$ is smaller than $f_{c}$. The flow of $\lambda_l$ then
has to be included into our consideration and the renormalization of
$\tilde{f}_l$ will be found to stop at a larger scale.

Let us then analyze the flow of the correlator ${\tilde \Delta}_l$,
Eq.~(\ref{eq:RGcreep1}), at finite temperatures in more detail. The
non-localities introduced by the finite value of $\lambda_l$ in
(\ref{eq:RGcreep1}) can be neglected as long as $\lambda_l$ is smaller than the
length scale introduced by the finite temperature, and we can therefore continue
to use the quasi-static equation Eq.~(\ref{static}). Below, we will make use of
the flow equations only in the regime where this condition holds. We also
neglect the disorder contribution to the temperature renormalization in
(\ref{eq:RGcreep3}) since it does not influence the main result to the accuracy
desired here.

At finite but small $T > 0$ the correlator flow below $l_c$ does not differ much
from the zero temperature case. However, at $l_c$ no cusp occurs at the origin
--- rather, the correlator remains rounded on a characteristic scale $u_l^T$.
Assuming that outside the thermally dominated region close to the origin the
correlator has approached its zero temperature fixed point shape ${\tilde
\Delta}^*(u)$ we may estimate $u_l^T$ from Eq.~(\ref{static}) by equating the
third and fourth terms on the RHS,
\begin{equation}
\label{uT}
  u^T_l\approx \frac{\tilde{T}_l}{|\tilde{\Delta}^{*\prime}(0+)|}
  \approx \frac{A_d \Lambda^{d-2} T e^{-\theta l}}
               {\epsilon c \xi e^{-\zeta l_c}}
  \approx \xi^*\frac{T}{U_c}e^{-\theta(l-l_c)},
\end{equation}
with $U_c$ the typical elastic energy on the Larkin scale $L_c$, $U_c\approx
(\epsilon/A_d)\,c\xi^2 L_c^{d-2}$. Obviously, for low temperatures
$T \ll U_c$, the thermal rounding of the cusp involves a scale $u^T_l$
much smaller than the width $\xi^*$ of the correlator.
\begin{figure}
\centerline{\epsfxsize=8.0cm \epsfbox{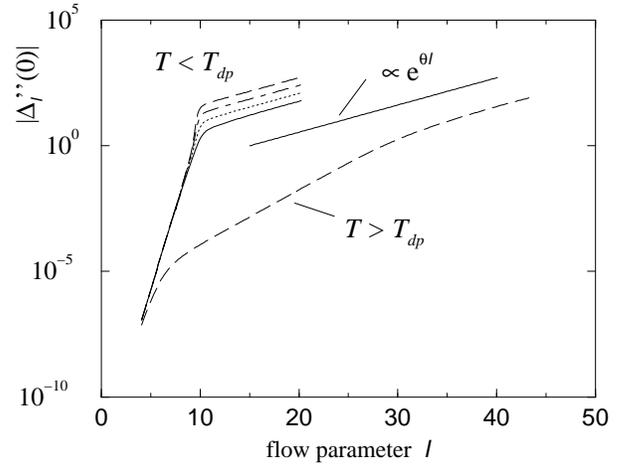}} \narrowtext\vspace{4mm}
\caption{Numerical integration of the curvature $|{\tilde\Delta}^{\prime
\prime}_l(0)|$ in the force correlator for $d=1$, see Eq.~(\ref{static}). Four
initial temperatures $T_0 < T_{\rm dp}$ below thermal depinning at $T_{\rm dp} =
U_c = c \xi^2 /L_c$ with values $T_0 = T_{\rm dp}/\alpha$, $\alpha = 40, 20, 10,
5$ have been chosen, while the high temperature curve starts with $T_0 = 20
T_{\rm dp}$. The initial growth through the Larkin regime involves the exponents
$\epsilon = 3$ and $\epsilon - 5 \zeta_{\rm th} = 1/2$ at low and high
temperatures, respectively. Beyond the Larkin regime the flow of
${\tilde\Delta}^{\prime \prime}_{l}(0) \propto -1/{\tilde T}_{l}$ is
characterized by the temperature exponent $\theta = 2\zeta -1 = 0.2495$ (c.f.,
(\ref{curvature}) with a wandering exponent $\zeta$ given by the one-loop
fixed-point value\cite{DSFisher} $\zeta \approx 0.2083\epsilon$). The crossover
at the Larkin scale is sharp and independent of temperature below $T_{\rm dp}$;
the uniform vertical spacing of the asymptotic curves reflects the temperature
independence of the energy scale $U_c$. For high temperatures above $T_{\rm dp}$
the crossover is shifted to $l_c(T) \gg l_c$ and the pinning energy $U_c(T)$
depends on temperature, $U_c \approx T$ (note that $|{\tilde\Delta}^{\prime
\prime}_{l_c(T)}(0)| \sim 1$).} \label{F2}
\end{figure}
The finite temperature curvature of the correlator follows from a comparison of
the term $\tilde{T}_l \tilde{\Delta}^{\prime\prime}_l$ with either the fourth or
the last term in (\ref{static})
\begin{equation}
 {\tilde\Delta}^{\prime\prime}_{l\gg l_{c}}(0)
 \approx -\frac{{{\tilde\Delta}^{*\prime}}(0+)^{2}}{{\tilde T}_{l}}
 \simeq -\frac{U_{c}}{T}e^{\theta (l-l_{c})};
 \label{curvature}
\end{equation}
the finite temperature fluctuations thus regularize the `cuspy' divergence
occurring at $T=0$. The above estimates are correct up to numerical prefactors
only; a more rigorous derivation can be found in Ref.\ \onlinecite{Chauve}.

For the following it is crucial to establish that the behavior described by
Eq.~(\ref{curvature}) is valid already soon after $l_c$, as the flow equation
(\ref{eq:RGcreep2}) for $\lambda_l$ is quite sensitive to the curvature of the
correlator at $u=0$. Indeed, as we have checked numerically (see Fig.~\ref{F2}),
after a rapid growth within the Larkin regime, the curvature
$\tilde{\Delta}^{\prime\prime}_l(0)$ saturates at a value $\sim U_c/T$ with a
slow further growth due to temperature rescaling, ${\tilde
\Delta}^{\prime\prime}_l(0)\propto\tilde{T}_l^{-1}\propto e^{\theta l}$. As long
as we are interested in the threshold behavior of the barrier close to $f_c$ it
is sufficient to establish a rapid crossover of the curvature from a steep
growth below $l_c$ to a gentle increase above $l_c$ (which we will neglect for
small $l-l_c>0$). Also note that the crossover occurs essentially at the same
value of $L_c$ independent of the temperature $T$, while the uniform vertical
spacing of the asymptotic curves reflects the temperature independence of the
energy scale $U_c$.

Let us now analyze the system of equations (\ref{eq:RGcreep1}) --
(\ref{eq:RGcreep3}) for $T > 0$, $v > 0$ close to criticality $f < f_{c}$. We
assume a fixed deviation $f_{c}-f$ from the threshold and a small temperature $T
\ll U_c$. The velocity then is (exponentially) small as well and represents the
smallest parameter in the problem. In the opposite case ($T$ small and fixed
while $f_{c}-f\rightarrow 0+$) the velocity $v \propto T^{1/\delta}$ has been
argued to scale as a power of temperature \cite{Middleton1,Fisher85}, the
numerical value of the exponent $\delta$ still being a matter of controversy
\cite{Middleton1,Nowak,Usadel}.

Again, we consider separately the two regimes below and above $l_c$. Throughout
the Larkin regime the parameter $\lambda_l$ remains small and can be set to zero
in (\ref{eq:RGcreep1}) and (\ref{eq:RGcreep4}). Furthermore, the temperature
dependent term in (\ref{static}) may be neglected initially --- it will become
relevant when ${\tilde\Delta}_{l}^{\prime\prime}(0)$ becomes of the order of
$U_{c}/T$. The flow of $\lambda_l$ through the Larkin regime then follows from
expanding the flow equations for ${\tilde\Delta}_{l}^{\prime\prime}(0)$ and
$\lambda_l$, (\ref{eq:RGcreep1}) and (\ref{eq:RGcreep2}), to second order in the
small parameter $\lambda_l$ and setting $\tilde{T}_l=0$,
\begin{equation}
 \label{eq:renlambda}
  \partial_l(\ln |\tilde{\Delta}_l^{\prime\prime}(0)|-3\ln\lambda_l)
  = \epsilon-3(2-\zeta)+{\mathcal O}(\lambda_l^4).
 \end{equation}
Integrating from $0$ to $l_c$ and using $\tilde{\Delta}_{l_c}^{\prime\prime} (0)
\approx U_c/T$ we obtain (up to numerical factors)
\begin{equation}
  \label{lambdalc}
  \lambda_{l_{c}} \simeq \frac{\eta v L_c^2}{c} e^{-\zeta l_{c}}
  {\left (\frac{U_{c}}{T}\right )}^{{1}/{3}}.
\end{equation}
Note that $\lambda_{l_c}$ grows as a {\it power} of $U_c/T$ as the temperature
approaches zero, whereas in the following depinning regime $\lambda_l$ will be
{\it exponentially} sensitive to $T$.

Going beyond the Larkin regime $l > l_c$ the function ${\tilde\Delta}_{l}(u)$
quickly approaches its fixed point form except for a small thermally smoothed
region of size $u^T_{l}$ around the origin with the second derivative given by
Eq.~(\ref{curvature}). As long as $\lambda_l < u_l^T$ one can keep $\lambda_l=0$
in the integral on the RHS of Eq.~(\ref{eq:RGcreep2}) and a simple integration
from $l_c$ to $l$ provides the result
\begin{eqnarray}
 \lambda_{l}
 &\simeq& \lambda_{l_{c}}
  \exp\Bigl[(2-\zeta )(l-l_c)+\frac{U_{c}}{\theta T}
  \bigl(e^{\theta (l-l_{c})}-1\bigr)
  \Bigr]\nonumber \\
 &\sim& \lambda_{l_c} \exp\Bigl[
 \frac{U_{c}}{T}(l-l_{c})\Bigr],
  \label{eq:lambda}
 \end{eqnarray}
where in the last step we have assumed that $T/U_c \ll l-l_c \ll 1$.

Turning next to the force equation (\ref{eq:RGcreep4}) we note that at finite
temperature the disorder contribution adds in only at a larger scale $l_d > l_c$
where $\lambda_{l}$ becomes of the order of the thermal rounding $u^T_{l}$, in
contrast to the zero temperature case where disorder jumps in at $l_c$. The
condition
\begin{equation}
  \lambda_{l_{d}}\simeq u^T_{l_d}\simeq
  {\tilde T}_{l_{d}}/{|{\tilde\Delta}^{*\prime}(0+)|}
\label{condition}
\end{equation}
then determines a relation connecting the crossover scale $l_d$ with the initial
velocity $v$,
\begin{equation}
  v \sim \frac{f_c}{\eta}\left(\frac{T}{U_c}\right)^{4/3}
  \exp\left[-\frac{U_c}{T}(l_d-l_c)\right].
  \label{vld}
\end{equation}
Below $l_d$ the flow of ${\tilde f}_{l}$ is determined by the scaling term
alone, $\partial_{l}{\tilde f}_{l}=(2-\zeta ){\tilde f}_{l}$, and a simple
integration gives $\tilde f_l = f \exp[(2-\zeta) l]$, where we have dropped the
small correction $\eta v$ in the definition of $\tilde{f}$, $\tilde{f}=f-\eta
v$. At $l_d$ the disorder correction turns on rapidly and we enter the depinning
regime\cite{Chauve} at scales $l > l_d$. In this regime we can substitute the
fixed point correlator $\tilde{\Delta}^{*}$ for $\tilde{\Delta}_l$ since now
$\lambda_l \gg u^T_l$. Furthermore, since still $\lambda_l \ll \xi^{*}$ we can
set $\lambda_{l}$ equal to $0+$ in Eq.~(\ref{eq:RGcreep4}) and we obtain a
disorder correction $c\Lambda^2\tilde{\Delta}^{*\prime}(0+)$. As argued before
when determining the threshold force $f_c$, we have to prevent the force $f_l$
from running away to $\pm\infty$ and thus the disorder term has to match the
scaling term; we then arrive at a second relation expressing $l_d$ in terms of
the applied driving force $f$,
\begin{equation}
 f e^{(2-\zeta) l_{d}}
 \approx \frac{c\Lambda^{2}|{\tilde\Delta}^{*\prime}(0+)|}{2-\zeta}
 \equiv f_{c}e^{(2-\zeta )l_{c}}.
\label{fld}
\end{equation}
With $l_{d}$ close to $l_{c}$ we can expand, $l_{d}-l_{c}\simeq (2-\zeta)^{-1}\,
(1-f/f_{c})\ll 1$ and combining with (\ref{vld}) we arrive at the final result
for the average velocity $v$
\begin{equation}
  v \propto \exp\left\{-\frac{U_{c}}{T}\frac{f_{c}-f}{f_{c}}\right\},
  \label{final}
\end{equation}
where we have dropped an unessential numerical factor in a redefinition of
$U_c$. Also, our analysis is not sufficiently precise to specify the prefactor.

\begin{figure}
\centerline{\epsfxsize=7.5cm \epsfbox{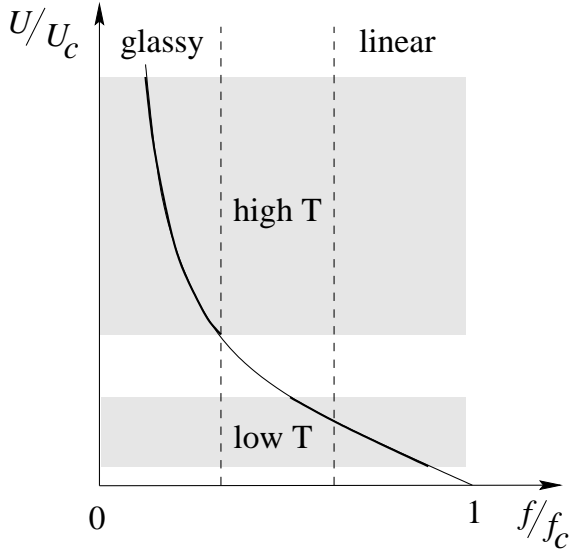}} \narrowtext\vspace{4mm}
\caption{Effective creep barrier $U$ at low temperatures as a function of
external force $f$. The thin line follows the interpolation formula $U(f) \simeq
U_c[(f_c/f)^\mu-1]$, properly interpolating between the glassy and linear
regimes at small drives and close to threshold, respectively. The slow
relaxation governed by the logarithmic decay law $U(f) \approx -T \ln(1+t/t_0)$
limits the experimental window to the interval $5 < U/T < 30$ depending on
temperature. Typically, a low (high) temperature measurement then probes the
linear (glassy or non-linear) regime.} \label{F3}
\end{figure}

Summarizing, we find that close to threshold with $T/U_c \ll 1-f/f_c <
2/(1+\mu)$ the velocity obeys an Arrhenius-type law with an energy barrier
decreasing linearly on approaching $f_c$. On the other hand, the usual glassy
behavior $U(f) \sim U_c (f_c/f)^\mu$ is valid at small forces $f/f_c <
2^{-1/\mu}$. In typical experiments the measured barriers are related to the
waiting time $t$ in the experiment, $U(f) \sim T \ln(t/t_0)$, and only a limited
regime of forces with barriers $5 < U(f)/T < 30$ is available. This regime is,
by making use of an extended temperature interval, still sufficient to probe
both the linear and the glassy regimes close to threshold and at low drives,
respectively, see Fig.~\ref{F3}.

\section{Creep at High Temperatures}\label{temperaturecreep}

In this section we consider thermal creep of $d=1,\,2$ dimensional elastic
interfaces (strings and interfaces moving in one transverse direction) at high
temperatures and for small driving forces. We show that the velocity-force
characteristic exhibits a non-Arrhenius type behavior $v \propto \exp
\left[-{\left ({f_{c}(T)}/{f}\right )}^{\mu}\right]$. This dependence derives
from a creep-type motion with a renormalized activation barrier of the order of
temperature, $U_c(T) \sim T$, and involves only the renormalized critical force
$f_c(T)$. For $d=3$ the parameters $f_c(T)$ and $U_c(T) \gg T$ depend on
temperature as well, leading to a non-Arrhenius type creep which is
non-universal, however, with a result depending explicitly on the chosen cutoff.
In Section \ref{creep1} we show how to calculate the renormalized energy barrier
$U_c(T)$ and the thres\-hold force density $f_c(T)$ using scaling arguments and
then present a more rigorous analysis using (dynamical) FRG in Sections
\ref{creep2} and \ref{creep3}.

\subsection{Scaling Analysis}\label{creep1}

While the mean squared displacement $\langle u^2 \rangle_{\rm th} \approx T
\Lambda/\pi c$ is bounded in $d=3$ (with $\Lambda^{-1}$ the intrinsic cutoff of
the manifold), the thermal displacement (or wandering) of strings and interfaces
($d=1,\,2$) grows unboundedly, either with distance $z$ or time $t$,
\begin{eqnarray}
\label{mus}
 \langle u^2({\bf z},t) \rangle_{\rm th} &\equiv&
 \langle [u({\bf z},t) - u({\bf 0},0)]^2 \rangle_{\rm th} \\
&=&
 \frac{2T}{c}\int \frac{d^dq}{(2\pi)^2} q^{-2}
 {\rm Re}\left[1-e^{i{\bf q}\cdot{\bf z}}e^{-(c/\eta)q^2 t}\right] \nonumber \\
&=&
 \left\{ \begin{array}{r@{\quad\quad}l}
 \displaystyle{\frac{T}{\pi c}\big[z^2+(c/\eta)t\big]^{1/2}}, & {d=1,}\\
 \noalign{\vskip 5 pt}
 \displaystyle{\frac{T}{2\pi c}\ln\big[\Lambda^2(z^2+(c/\eta)t)\big]}, &
                      {d=2;}\nonumber\\
  \end{array}\right.
\end{eqnarray}
the pinning length $L_c(T)$ set by the disorder landscape then has to provide
the necessary cutoff which in $d=3$ is given by the intrinsic cutoff
$\Lambda^{-1}$ (note that the $q$-integral in (\ref{mus}) is dominated by small
(large) $q$ for low (high) dimensions; hence, depending on the dimensionality of
the system the amplitude of thermal fluctuations is determined by short ($d=3$)
or long ($d=1,\, 2$) scales). This implies the existence or absence of a
separation of scales for thermal and disorder effects: While for $d=3$ these
scales are separated, $\Lambda^{-1} \ll L_c(T)$, no such separation is effective
in dimensions $d=1,\, 2$; thermal effects smearing the disorder landscape are
active on scales $L < L_c(T)$ while disorder takes over for $L > L_c(T)$, i.e.,
at the same scale. The temperature induced smoothing of the disorder potential
then follows different rules in low and high dimensions, as we are going to
discuss now.

In order to discuss pinning and creep we have to determine the renormalized
disorder landscape. Assuming pinning to involve longer time scales than thermal
fluctuations we average the pinning potential over thermal
excursions\cite{Blatter},
\begin{eqnarray}
\label{Epin}
 \langle\langle E_{\rm pin}^2 (L) \rangle_{\rm t}\rangle &=&
 \int_0^{t_0}\frac{dt}{t_0}\int_0^{t_0}\frac{dt'}{t_0}\int d^dz \int d^dz'
 \nonumber \\ && \qquad \times
 \langle V(u({\bf z},t),{\bf z})V(u({\bf z'},t'),{\bf z'})\rangle \nonumber \\
 &=& L^d\int_0^{t_0}\frac{dt}{t_0}\int_0^{t_0}\frac{dt'}{t_0} K[|u(t)-u(t')|]
 \nonumber \\
 &\approx& L^d K(0)\left(\frac{\xi^2}{\langle u^2(\Lambda^{-1},L)\rangle_{\rm th}}
             \right)^{1/2},
\end{eqnarray}
where the mean squared thermal fluctuations $\langle u^2(\Lambda^{-1},L)
\rangle_{\rm th}$ are cutoff by $\Lambda^{-1}$ or $L=L_c(T)$ in high and low
dimensions, respectively. Here, $K(u)$ denotes the potential correlator which is
related to the force correlator $\Delta(u)$ used above via $-K^{\prime\prime}(u)
= \Delta(u)$. The result (\ref{Epin}) tells us that at high temperatures,
thermal fluctuations replace the basic length scale $\xi$ of the disorder
landscape by the scale $\langle u^2(\Lambda^{-1},L)\rangle_{\rm th}^{1/2} > \xi$
(note that the energy scale of the disorder potential remains unchanged).
Comparing this smoothed pinning energy with the elastic energy $c\langle
u^2(\Lambda^{-1},L)\rangle_{\rm th} L^{d-2} \sim T$ we obtain the new pinning
scale replacing the $T=0$ Larkin length $L_c = (c^2\xi^4/K(0))^{1/(4-d)}$,
\begin{eqnarray}
L_c(T) \sim L_c \left( \frac{c^2 \langle u^2(\Lambda^{-1},L)\rangle_{\rm
th}^{5/2}}{K(0)\xi} \right)^{1/(4-d)}. \label{lct}
\end{eqnarray}
For high dimensions we find $L_c(T) \sim L_c (T/T_{\rm dp})
^{1/2\xi_{\scriptscriptstyle F}}$ with the Flory exponent
$\xi_{\scriptscriptstyle F} = (4-d)/5$ and the depinning temperature $T_{\rm dp}
= c\xi^2/\Lambda$ (this result follows from simple scaling $u \propto
L^{\xi_{\scriptscriptstyle F}}$ and $u \propto T^{1/2}$). In dimensions $d=1,\,
(2)$ the corresponding result takes the form (here we concentrate on the case
$n=1$, see Ref.\ \onlinecite{Gorokhov} for a discussion of the marginal
situation in $d+n = 1+2$)
\begin{eqnarray}
L_c(T) \sim L_c \left(\frac{T}{T_{\rm dp}} \right)^{\lambda}, \label{lctld}
\end{eqnarray}
with the temperature exponent $\lambda = 5,\,(5/4)$ and the depinning
temperature $T_{\rm dp} = (cK(0)\xi^2)^{1/3},\,(c\xi^2)$. Comparing with the
above Flory exponent we see that thermal fluctuations indeed are much more
important in $d=1$, while for $d=2$ the corrections are only logarithmic (not
shown in (\ref{lctld})).

The energy barrier and the threshold force are renormalized correspondingly; for
$d=3$ we have $U_c(T) \sim U_c (T/T_{\rm dp})^{7/2} \gg T$, with $U_c = c \xi^2
L_c$ and $T_{\rm dp} = c\xi^2/\Lambda$, while for $d=1,\, 2$ the barrier
`saturates' above $T_{\rm dp}$, $U_c(T) \sim T$. The critical force density is
renormalized according to $f_c(T) \sim f_c (T_{\rm dp}/T)^{9/2}$ in $d=3$ and
takes the form
\begin{eqnarray}
f_c(T) \sim c\frac{\langle u^2(L_c(T))\rangle_{\rm
    th}^{1/2}}{L_c(T)^2}
\sim f_c \left(\frac{T_{\rm dp}}{T} \right)^{\kappa}, \label{fct}
\end{eqnarray}
with the temperature exponent $\kappa = 7,\,(2)$ in dimensions $d=1,\,
(2)$, restricting ourselves again to the case $n=1$.  Using the usual
creep formula $U(f)/T = (U_c/T) (f_c/f)^\mu$ and inserting the new
temperature dependent values for $U_c \rightarrow T$ and $f_c
\rightarrow f_c(T)$ we obtain the creep exponent $(f_c/f)^\mu (T_{\rm
  dp}/T)^{\kappa\mu}$ producing a non-Arrhenius type creep in $d=1,\,
2$ involving only the renormalized critical force density $f_c(T)$. For higher
dimensions $d > 2$ the renormalized barrier $U_c(T) \gg T$ remains large, a
consequence of the separation of scales mentioned above. In the following we
rederive these scaling results via the more rigorous analysis provided by the
(dynamical) FRG scheme.

\subsection{Functional Renormalization Group}\label{creep2}

In a first step we rederive the crossover scale $L_c(T)$ via the functional
renormalization group. The nonlinear terms in the flow equations are still small
during the initial stage of the RG flow --- neglecting them we first solve the
linear equation. The length $L_{c}(T)$ then appears as the characteristic length
where the nonlinear corrections become of the order of the linear terms. The
analysis is conveniently carried out for the potential correlator ${\tilde
K}_{l}(u)$ which follows the flow equation
\begin{eqnarray}
  \partial_{l}{\tilde K}_{l}(u)
  &=& (\epsilon-4\zeta ){\tilde K}_{l}(u)+\zeta u {\tilde K}_{l}^{\prime}(u)
  +{\tilde T}_{l}{\tilde K}_{l}^{\prime\prime}(u)\nonumber \\
  &&\qquad
  +\frac{1}{2}{\tilde K}_{l}^{\prime\prime}(u)^{2}
  -{\tilde K}_{l}^{\prime\prime}(u){\tilde K}_{l}^{\prime\prime}(0),
  \label{frgk}
  \end{eqnarray}
while the temperature flow is given by
\begin{equation}
  \partial_{l}{\tilde T}_{l}=-\theta{\tilde T}_{l}.
  \label{Tflow}
  \end{equation}
Using the ansatz
\begin{equation}
  \label{P}
  {\tilde K}_l(u)=\exp\left[(\epsilon-4\zeta-(2-d)/2)l\right]
  {\tilde P}_l(ue^{\theta l/2})
\end{equation}
the linear part of the flow transforms into a Fokker-Planck equation
describing the probability distribution ${\tilde P}_l(u)$ for an overdamped
particle moving in a parabolic potential at constant temperature
${\tilde T}_0$ \cite{Gorokhov},
\begin{equation}
  \label{FokkerPlanck}
  \partial_l {\tilde P}_l(u)=\frac{2-d}{2}\partial_u[u {\tilde P}_l(u)]
  +{\tilde T}_0\partial_u^2 {\tilde P}_l(u),
\end{equation}
for which the fundamental solution is well-known. Solving the initial value
problem for ${\tilde P}_l$ and inserting in Eq. (\ref{P}), we obtain
\begin{eqnarray}
  {\tilde K}_l(u)&=&\exp\left[(4-d-5\zeta)l\right]\Big[\frac{2-d}
  {4\pi{\tilde T}_0e^{-\theta l}(1\!-\!e^{-(2-d)l})}\Big]^{1/2}\nonumber\\
 && \int\! du' \, {\tilde K}_0(u')
    \exp\Big[\!-\!\frac{(2\!-\!d)(u\!-\!u'e^{-\zeta l})^2}
     {4{\tilde T}_0 e^{-\theta l}(1\!-\!e^{-(2-d)l})}\Big].\nonumber
\end{eqnarray}
For high temperatures such that ${\tilde T}_0 e^{-\theta l}(1-e^{-(2-d)l})
/(2-d) > \xi^2 e^{-2\zeta l}$ the factor ${\tilde K}_0(u')$ acts as a
$\delta$-function and can be extracted from the integral,
\begin{equation}
  \frac{{\tilde K}_{l}(u)}{\int {\tilde K}_{0}(u)\thinspace du} \approx
  \left\{ \begin{array}{r@{\quad\quad}l}
  \displaystyle{
  e^{(3-5\zeta)l}\frac{\exp\left[-\frac{u^{2}}{4{\tilde T}_{0}
  e^{-\theta l}}\right]}
  {(4\pi {\tilde T_{0}}e^{-\theta l})^{1/2}}},
  & {d=1,}\\ \noalign{\vskip 5 pt}
  \displaystyle{
  e^{(2-5\zeta)l}\frac{\exp\left[-\frac{u^{2}}{4{\tilde T}_{0}l
  e^{-\theta l}}\right]}
  {(4\pi {\tilde T_{0}}le^{-\theta l})^{1/2}},}
  & {d=2,}\\ \noalign{\vskip 5 pt}
  \displaystyle{
  e^{(1-5\zeta)l}\frac{\exp\left[-\frac{u^{2}}
  {4{\tilde T}_{0}e^{-2\zeta l}}\right]}
  {(4\pi {\tilde T_{0}}e^{-2\zeta l})^{1/2}},}
  & {d=3.}\\ \noalign{\vskip 5 pt}
\end{array}\right.
\label{otwet*}
\end{equation}
The above constraint on the temperature simplifies to ${\tilde T}_0 > e^{-l}
\xi^2$, ${\tilde T}_0 > \xi^2/l$, and ${\tilde T}_0 > \xi^2$ for
$d=1,\,2,\,3$, respectively, where $\xi$ is the initial width of the correlator
${\tilde K}_0$; expressing ${\tilde T}_0$ and $l$ through the physical
quantities $T$ and $L$ we recover the condition $\langle
u^2(\Lambda^{-1},L)\rangle_{\rm th} > \xi^2$.

The solutions (\ref{otwet*}) still involve the wandering exponent $\zeta$.
Although the final physical results do not depend on the particular choice, it
is a matter of convenience to adopt the thermal values $\zeta_{\rm
th}=1/2,\,\theta_{\rm th} = 0$ for $d=1$, $\zeta_{\rm th}=0,\,\theta_{\rm th}=0$
for $d=2$, and $\zeta_{\rm th}=0,\,\theta_{\rm th}=1$ for $d=3$ and have the
correlator flow towards a thermal fixed point. The renormalized correlator
(\ref{otwet*}) then behaves very differently for large and small dimensions: In
$d=3$ the transverse scale $u$ does not change (as we chose $\zeta = 0$) and the
initial correlator of width $\xi$ is replaced with a new correlator of width
$\langle u^2\rangle_{\rm th} \sim {\tilde T}_0 = T\Lambda/c$.
This contrasts with the situation
in $d=1$ where $u$ does rescale (as we chose $\zeta = 1/2$) and the physical
width of the correlator increases with $l$ to follow the mean thermal
displacement amplitude $\langle u^2(L)\rangle_{\rm th} \sim TL/c$. For $d=2$ the
physical width grows only logarithmically.

The flow (\ref{otwet*}) indicates that the thermal fixed point is
unstable as the amplitude of the disorder grows exponentially under
the FRG transformation. As the nonlinear terms in (\ref{frgk}) become
large beyond the scale $L_c(T)=\Lambda^{-1}e^{l_c(T)}$ we cannot neglect
them any longer and the flow crosses over to approach the disorder
dominated fixed point (the wandering exponent $\zeta$ then has to be modified
accordingly, $\zeta=0.2083\epsilon$ for random bond disorder \cite{DSFisher}).
The pinning length $L_c(T)$ replaces the $T=0$ Larkin length $L_c$ and
can be found from a comparison of linear and quadratic terms
in the flow equation (\ref{frgk}),
\begin{equation}
\label{crosscond}
  {\tilde K}_{l_c(T)}(0)\simeq {\tilde K}^{\prime\prime}_{l_c(T)}(0)^2,
\end{equation}
making use of the result (\ref{otwet*}). It is easily verified that the
crossover condition (\ref{crosscond}) together with the explicit solution
(\ref{otwet*}) of the linearized flow equations then yields the results
(\ref{lct}) for the crossover length $L_c(T)$ obtained above with the help of
scaling arguments. Note that for $d=1,\,2$ the linear term ${\tilde T}_l{\tilde
K}^{\prime\prime}_{l}$ in (\ref{frgk}) gains in importance as we integrate
through the Larkin regime, hence thermal rounding persists on all scales $l <
l_c(T)$. On the contrary, for $d=3$ the thermal rounding term is most important
at small scales $l \sim 1$ where it quickly replaces the width $\xi$ of the
correlator by the mean thermal displacement amplitude $\langle u^2 \rangle_{\rm
th}^{1/2}$; upon further scaling the effective temperature decreases and at the
Larkin scale the thermal term is down by a factor of $e^{-l_c(T)}$. This again
reflects the different role the temperature plays for different internal
dimensions of the elastic manifold.

\subsection{Dynamic Functional Renormalization Group}\label{creep3}

Finally, let us see how the high temperature creep (as characterized by the
temperature dependent critical force density $f_c(T)$ for $d=1,\,2$) appears
directly from the dynamic FRG treatment. The analysis parallels the treatment at
low temperatures, however, we have to be more careful in distinguishing between
the cases $d=1,\,2$ and $d=3$.

Let us analyze the flow of the force correlator ${\tilde \Delta}_l(u) = -{\tilde
K}_{l}^{\prime\prime}(u)$. Following the full flow up to $l_c(T)$ the force
correlator assumes a shape with a height and width as given by (\ref{otwet*}).
At $l_c(T)$ the non-linear terms in the flow equation (\ref{frgk}) have become
important; beyond $l_c(T)$ the correlator quickly flows towards the disorder
dominated fixed point function ${\tilde \Delta}^\ast(u)$, the linear temperature
term ${\tilde T}_{l}{\tilde \Delta}_{l}^{\prime\prime}(u)$ smoothing the flow in
a region of size $u_l^{T}$ around the origin. Assuming again that the fixed
point function ${\tilde \Delta}^\ast(u)$ derives is rough shape from the
correlator ${\tilde \Delta}_{l_c(T)}(u)$ at crossover, we can use the result
(\ref{otwet*}) in combination with the flow equation (\ref{static}) to find the
characteristic features ${\tilde \Delta}^*(0)$, $\xi^*$, and ${\tilde
\Delta}^{*\prime}(0+)$ of the fixed point function and the rounding parameters
${\tilde \Delta}_l^{\prime\prime}(0)$ and $u_l^{T}$ of the cusp.

Using (\ref{otwet*}) and the crossover condition (\ref{crosscond}) we find the
height ${\tilde \Delta}^*(0) = -{\tilde K}_{l_c(T)}^{\prime\prime}(0) \approx
{\tilde T}_0$. The slope of the fixed point function at $u=0+$ again follows
from comparing terms in the flow equation, $|{\tilde \Delta}^{*\prime} (0+)|
\approx {\tilde \Delta}^*(0)^{1/2}$ $\approx ({\tilde T}_{0})^{1/2}$ and we find
the width $\xi^* \approx ({\tilde T}_{0})^{1/2}$ $\approx \langle u^2(L_c(T))
\rangle_{\rm th}^{1/2} e^{-\zeta_{\rm th} l_c(T)}$, see also the result
(\ref{otwet*}). The width $u_l^T$ of thermal rounding derives from $u_l^T
\approx {\tilde T}_l/|{\tilde \Delta}^{*\prime}(0+)|$ and we find the result
$u_l^T/\xi^* \sim e^{-\theta(l-l_c(T))}$ in $d=1,\,2$, with $\theta = d-2+2
\zeta$ and $\zeta$ the random mani\-fold exponent; thus at $l_c(T)$ the width of
thermal rounding equals the width of the correlator, $u_l^T \approx \xi^*$. The
curvature ${\tilde \Delta}_l^{\prime\prime}(0) \sim {\tilde \Delta}^*(0)/{\tilde
T}_l$ is correspondingly small, ${\tilde \Delta}_l^{\prime\prime}(0) \approx
-e^{(l-l_c(T))}$, see Fig.\ 2; comparing this result with Eq.\ (\ref{curvature})
we conclude that the barriers `saturate' to follow the temperature, $U_c(T)
\approx T$. This is quite different from $d=3$: Here, the thermal rounding
affects only the narrow regime $u_l^T \approx \xi^* e^{-l_c(T)} e^{-\theta
(l-l_c(T))} \approx (U_c(T)/T)\, e^{\theta(l-l_c(T))}$ around the origin and the
curvature is already large at $l_c(T)$, ${\tilde \Delta}_l^{\prime\prime}(0)
\approx -e^{l_c(T)} e^{\theta(l-l_c(T))} = -(U_c(T)/T)\, e^{\theta(l-l_c(T))}$,
where $U_c(T) = c\langle u^2\rangle_{\rm th} L_c(T) \gg T$. Thus, as anticipated
above, the curvature of the correlator at $l_c(T)$ is thermally reduced to the
order of $1$ in $d=1,\,2$ while it is large in $d=3$. Hence for $d=3$ the
situation at high temperatures $T > T_{\rm dp}$ is not different from that at
low temperatures.

Next we integrate the flow for the velocity and force parameters $\lambda_l$ and
$f_l$. We determine the creep scale $l_d(T)$ twice, using the condition
$\lambda_{l_d(T)} = u^T_{l_d(T)}$ to relate the velocity $v$ and the scale
$l_d(T)$ and a second time from the onset of the disorder term in the force
equation (\ref{fld}), providing a relation between $f$ and $l_d(T)$; combining
these results we obtain the desired velocity-force characteristics. In doing so,
we have to be careful to use the above high temperature estimates for ${\tilde
\Delta}^{*\prime}(0+)$ in Eqs.\ (\ref{condition}) and (\ref{fld}).

Integrating the flow equation (\ref{eq:RGcreep2}) for $\lambda_l$ through the
Larkin regime and then up to $l_d(T)$ we find the first relation
\begin{equation}
v \propto \exp[-e^{\theta(l_d(T)-l_c(T))}]. \label{vldT}
\end{equation}
Integrating next the force equation (\ref{eq:RGcreep3}) we obtain
\begin{equation}
f_{l_d(T)} \sim f e^{(2-\zeta_{\rm th}) l_c(T) + (2-\zeta) [l_d(T)-l_c(T)]}
\label{fldT0}
\end{equation}
and equating this to the disorder induced term $c \Lambda^2
|\Delta^{*\prime}(0+)|$ in the flow equation (\ref{eq:RGcreep3}) we arrive at
the second relation
\begin{equation}
f \sim f_c(T) e^{-(2-\zeta)[l_d(T)-l_c(T)]} \label{fldT}
\end{equation}
with the critical force density $f_c(T) = c {\tilde T}_0^{1/2} \Lambda^2$
$e^{-(2-\zeta_{\rm th})l_c(T)}$ $\sim c \langle u^2(L_c(T)) \rangle_{\rm
th}^{1/2}/L_c^2(T)$, in agreement with (\ref{fct}). Combining the results
(\ref{vldT}) and (\ref{fldT}) we find the velocity-force characteristic
describing the non-Arrhenius type creep at high temperature,
\begin{eqnarray}
  \label{finalht}
   v \propto
   \left\{ \begin{array}{r@{\quad\quad}l}
      \displaystyle{
        \exp\left[-\left(\frac{f_c(T)}{f}\right)^{\theta/(2-\zeta)}\right]},
        & {d=1,\,2,} \\ \noalign{\vskip 5 pt}
      \displaystyle{\exp\left[-\frac{U_c(T)}{T}
                        \left(\frac{f_c(T)}{f}\right)^{\theta/(2-\zeta)}\right]},
        & {d=3;}\nonumber\\
  \end{array}\right.
\end{eqnarray}
In conclusion, using dynamical functional renormalization group theory we have
derived the linear scaling of the creep barriers close to $f_c$ and have put the
non-Arrhenius type high temperature creep of low-dimensional manifolds on a firm
basis. The simple behavior of the creep barrier close to threshold appears
surprising --- considering the `non-trivial' threshold exponents due to a
diverging nucleus obtained for elastic manifolds trapped in a washboard
potential (see Ref.\ \onlinecite{Blatter}) one is tempted to expect a
non-trivial exponent for the random case as well. However, from our analysis we
conclude that there is no new diverging scale associated with creep near
threshold. The linear decay of the creep barrier then follows from a regular
expansion and no effects of critical fluctuations are picked up.

We thank Pascal Chauve for discussions and the Swiss National Foundation for
financial support.

\end{multicols}

\end{document}